\documentclass[twocolumn,superscriptaddress,showpacs,pra]{revtex4}
\DeclareUnicodeCharacter{2212}{\textendash}
\usepackage{epsfig}
\usepackage{epsfig}
\usepackage{epstopdf}
\usepackage{delarray}
\usepackage{amsmath, amssymb}
\usepackage{bm}
\usepackage{graphicx}
\usepackage{placeins}
\usepackage{color} 

\begin{document}

\title[Rogue quantum gravitational waves]{Rogue quantum gravitational waves}

\author{Cihan Bay\i nd\i r}
\affiliation{\.{I}stanbul Technical University, Engineering Faculty, 34469, Maslak, \.{I}stanbul, Turkey.\\  
Bo\u{g}azi\c{c}i University, Engineering Faculty, 34342 Bebek, \.{I}stanbul, Turkey. \\
International Collaboration Board Member, CERN, CH-1211 Geneva 23, Switzerland.}
\email{cbayindir@itu.edu.tr}

\author{Fatih Ozaydin}
\affiliation{Institute for International Strategy, Tokyo International University, 1-13-1 Matoba-kita, Kawagoe, Saitama, 350-1197, Japan.\\
International Collaboration Board Member, CERN, CH-1211 Geneva 23, Switzerland.}

\author{Azmi Ali Altıntaş}
\affiliation{Department of Physics, Faculty of Science, Istanbul University, 34116, Vezneciler, Istanbul, Turkey\\ 
International Collaboration Board Member, CERN, CH-1211 Geneva 23, Switzerland.}

\author{Metin Ar\i k}
\affiliation{Bo\u{g}azi\c{c}i University, Faculty of Arts and Sciences, 34342 Bebek, \.{I}stanbul, Turkey.}

\vspace{10pt}

\begin{abstract}
In this paper, we propose the existence and discuss the properties of rogue quantum gravitational waves. More specifically, we numerically solve the Schrödinger-Newton system of equations using a spectral scheme with a $4^{th}$ order Runge-Kutta time integrator and show that noise either imposed on wave function $\Psi$, or the gravitational field $\Phi$, triggers the modulation instability which turns the monochromatic wave fields into chaotic ones exhibiting high and unexpected waves. Such waves can be named as rogue quantum gravitational waves. We discuss the characteristics and probabilities of occurrences of such rogue waves in the frame of the Schrödinger-Netwon equations. We suggest alternative methods for studying rogue quantum gravitational waves and rogue gravitational waves.

\pacs{04.60.−m, 04.30.−w}
\end{abstract}
\maketitle

\section{Introduction}

Quantum gravity attempts to describe the quantum behavior of the gravitational field \cite{Kiefer, Misner, Feynman95, Feynman2003, Kenyon}. This field of science emerged due to two significant necessities. First of them is the unification of quantum field theory with the classical theory of general relativity \cite{Kiefer}. The second one is the understanding of the early Universe near Big Bang and the final stages of black-hole evolution \cite{Kiefer}. With these motivations, various models for quantum gravity have been proposed in the literature \cite{Kiefer, vanMeter}.

The Schrödinger-Newton (S-N) equation system is the restricted version of the semi-classical Einstein equations in the Newtonian limit \cite{Kiefer, Lange, Bahrami, Albers}. In the S-N equation system, the Schrödinger equation is coupled with a Newtonian gravitational potential, where the gravitational potential emerges due to the wave function behaving as a mass density \cite{Kiefer, Robertshaw, Moroz, TodMoroz}, thus this system includes a term that accounts for the interaction of a particle with its own gravitational field. The S-N equation system has been proposed by Diósi \cite{Diosi84} and Penrose \cite{Penrose96, Penrose98} for modeling the dynamics of the collapse of the wave function in quantum mechanics. In this model, the gravity is classical even at the fundamental level, however matter has quantum behavior, as extensively discussed in the literature \cite{Harrison, Illner, Ghirardi86, Ghirardi90, Diosi87, Diosi89}. Some of the studies on the S-N equation system include analysis of the existence and asymptotic behaviour of the 3D version of the system in the attractive case with positive energy \cite{Illner, Arriola}, its eigenstates \cite{Bernstein} and ground state energy \cite{Tod2001} and wavefunction localization and decoherence \cite{Diosi07}. Additionally, an S-N system with a complex Newton constant and induced gravity has been studied to avoid the divergence of the S-N system \cite{Diosi09}. Recently, the S-N equation system has been the focus of a quite a debate since it is connected with the quantization of the gravity \cite{Carlip} and since it may be falsifiable when envisaged experiments are designed \cite{YangMiao, Giulini1, Giulini2, Giulini3}. Additionally, how it relates to well-established principles of physics has also been questioned \cite{Anastopoulos}. Although some criticisms mentioned above are made, the S-N equation system remains as a commonly studied system in the field of quantum gravity as the literature summarized above indicates. A more comprehensive discussion of S-N equations and their foundation can be seen in the review paper \cite{Bahrami}.

Rogue waves, on the other hand, became an active area of research in the last few decades \cite{Peregrine, Kharif,  Akhmediev2009a, Akhmediev2009b, Akhmediev2011, dqiu}. For marine and optical engineering studies, the rogue waves are defined as the waves having a height of at least 2 times the significant wave height in a chaotic wavefield \cite{Peregrine, Kharif, Akhmediev2009a, Akhmediev2009b, Akhmediev2011}. As their name implies, rogue waves have suddenly changing and extreme behavior in the wavefield. They appear in hydrodynamic medium \cite{Kharif}, fiber optics and quantum studies \cite{Akhmediev2009b, Akhmediev2011, Bay_Zeno, Birkholz}, Bose-Einstein condensates \cite{Akhmediev2009a, Akhmediev2009b, Akhmediev2011}, just to name a few field. For the modeling of rogue waves generally the nonlinear Schrödinger equation (NLSE) \cite{Schrodinger}, or NLSE like equations \cite{Akhmediev2009a, Soto2014RwSSchaotic, BayPRE1, BayPRE2} are commonly utilized. Additionally, the exact fully nonlinear wave equations such as the one discussed in \cite{Baysci}, is also used to model such extreme waves in the ocean environment.

Existence of gravitational waves was first predicted by Henri Poincaré in 1905 and Albert Einstein in 1916. With the recent observation of gravitational waves by LIGO for the first time, gravitational waves are currently an active research area. However, to our best knowledge neither the existence of rogue gravitational waves or rogue quantum gravitational waves have been questioned up-to-date, nor their dynamics. This paper aims to address the latter problem. More specifically, we examine the unexpected and high amplitude fluctuations in the frame of the S-N equation system of quantum gravity. We show the modulation stability (MI) turns the monochromatic wave fields into chaotic ones, thus triggers the formation of unexpected and high amplitude fluctuations in the frame of the S-N equations. Such oscillations can be named as rogue quantum gravitational waves. We discuss the properties, dynamics and statistics of these rogue quantum gravitational waves in the frame of the S-N equation system. We discuss our findings, comment on our results, their usability and significance.

\section{Methodology}
The Schrödinger-Newton (S-N) coupled system of equations can be given as
\begin{equation}
\begin{split}
i \hbar \frac{\partial \Psi ({\bf x},t)}{\partial t}=-\frac{\hbar ^2 }{2 m}\nabla^2 \Psi ({\bf x},t)& +V ({\bf x},t) \Psi ({\bf x},t) \\
& +m \Phi ({\bf x},t) \Psi ({\bf x},t)
\end{split}
\label{eq01}
\end{equation}

\begin{equation}
\nabla^2 \Phi ({\bf x},t)=4 \pi Gm \left| \Psi ({\bf x},t) \right|^2 
\label{eq02}
\end{equation}
where $\Psi ({\bf x},t)$ is the complex wave function,  $\Phi ({\bf x},t)$ is the gravitational field, $\hbar$ is the reduced Planck's constant, $m$ is the particle mass, $G$ is the gravitational constant and $i$ is the imaginary unity. In here, $V ({\bf x},t)$ is the ordinary potential and we assume its absence throughout this paper for simplicity. We consider the dynamics of S-N equations in 1D, thus the Laplacian becomes $\nabla^2 =\partial^2 /\partial x^2$. We work in SI units, therefore the parameters of computations are selected as $\hbar \approx 1.054 \times  10^{-34} \ kg m^2 s^{-1}$, $G \approx 6.673 \times 10^{-11} \ kg^{-1}m^3 s^{-2}$. In order to investigate the rogue wave dynamics of a proton, the particle mass is selected as $m \approx 1.672 \times 10^{-27} \ kg$. Using these parameters, we solve this governing system of S-N equations numerically. Starting from the initial conditions, the time stepping of the S-N equation system are performed using a $4^{th}$ order Runge-Kutta scheme. With this motivation, we rewrite the Eq. (\ref{eq01}) as
\begin{equation}
\begin{split}
\frac{\partial \Psi ({\bf x},t)}{\partial t} & =  \frac{-i}{\hbar} \left( -\frac{\hbar ^2 }{2 m}\frac{\partial^2 \Psi}{\partial x^2} ({\bf x},t)+m \Phi ({\bf x},t) \Psi ({\bf x},t) \right) \\ 
& =r(x, t, \Psi, \Phi)
\end{split}
\label{eq03}
\end{equation}
where the function $r(x,t, \Psi, \Phi)$ refers to the right hand side of Eq.(\ref{eq03}). The four slopes of the $4^{th}$ order Runge-Kutta scheme can be calculated at each time using
\begin{equation}
\begin{split}
& s_1=r(x, t^n, \Psi^n, \Phi^n) \\
& s_2=r(x, t^n+0.5dt, \Psi^n+0.5 s_1dt, \Phi^n) \\
& s_3=r(x, t^n+0.5dt ,\Psi^n+0.5 s_2dt, \Phi^n) \\
& s_4=r(x, t^n+dt, \Psi^n+s_1dt, \Phi^n) \\
\end{split}
\label{eq04}
\end{equation}
where $n$ shows the iteration count and $dt$ is the time step which is selected as $dt=0.1$ throughout this study. At each time step, the Laplacian term is computed using Fourier series as
\begin{equation}
\frac{\partial^2 \Psi}{\partial x^2} = F^{-1} \left[-k^2F[\Psi] \right] 
\label{eq05}
\end{equation}
where $F$ and $F^{-1}$ show the Fourier and the inverse Fourier transforms, respectively, and $k$ is the wavenumber vector which has exact $N$ multiples of the fundamental wavenumber, $k_o=2 \pi/(L)$  \cite{trefethen, Canuto}. In here, $L$ is the length of the spatial domain which is selected to be $L=100$. All nonlinear products are computed in the physical space. Starting from the initial conditions $\Psi_0$ and $\Phi_0$, $\Phi_{xx}$ is computed by Eq.(\ref{eq02}) at each time step. In order to integrate this $\Phi_{xx}$ term to get the $\Phi$, the FFT transform matrix is constructed by using a Toeplitz matrix and matrix division algebra is used to calculate the $\Phi$, as described in \cite{trefethen}. Then the calculated $\Phi$ is used in Eq.(\ref{eq04}) to calculate the 4 slopes of the Runge-Kutta algorithm. Then at each time step, the time parameter and the wave function $\Psi$ are computed using 
\begin{equation}
\begin{split}
& \Psi^{n+1}=\Psi^{n}+(s_1+2s_2+2s_3+s_4)/6 + \epsilon \\
& t^{n+1}=t^n+dt\\
\end{split}
\label{eq06}
\end{equation}
The initial condition for wave function is selected as $\Psi_0=a_{\Psi i}e^{ix}$, where $a_{\Psi i}$ is a constant. In Eq.(\ref{eq06}), the $\epsilon$ term represents the noisy input, which is selected as $\epsilon=\beta a_{\Psi i} v$ where $\beta$ is a constant and $v$ is an uniformly distributed random number set having values in the interval of $v \in [-1,1]$. This form of the initial condition and the random noisy input triggers the MI, thus rogue quantum gravitational waves are formed within the frame of the NS equations as discussed in the next section. Additionally, at each time step, the power of the complex wave function $\Psi$ is calculated using the expression $P_\Psi=\int |\Psi|^2dx$. Similarly, the power of the gravitational field $\Phi$ is calculated using $P_\Phi=\int \Phi^2dx$. These power calculations are useful for assessing the divergence of the model equations studied.

\section{Results and Discussion}

\begin{figure}[htb!]
\begin{center}
\hspace*{-0.7cm}
   \includegraphics[width=4.0in]{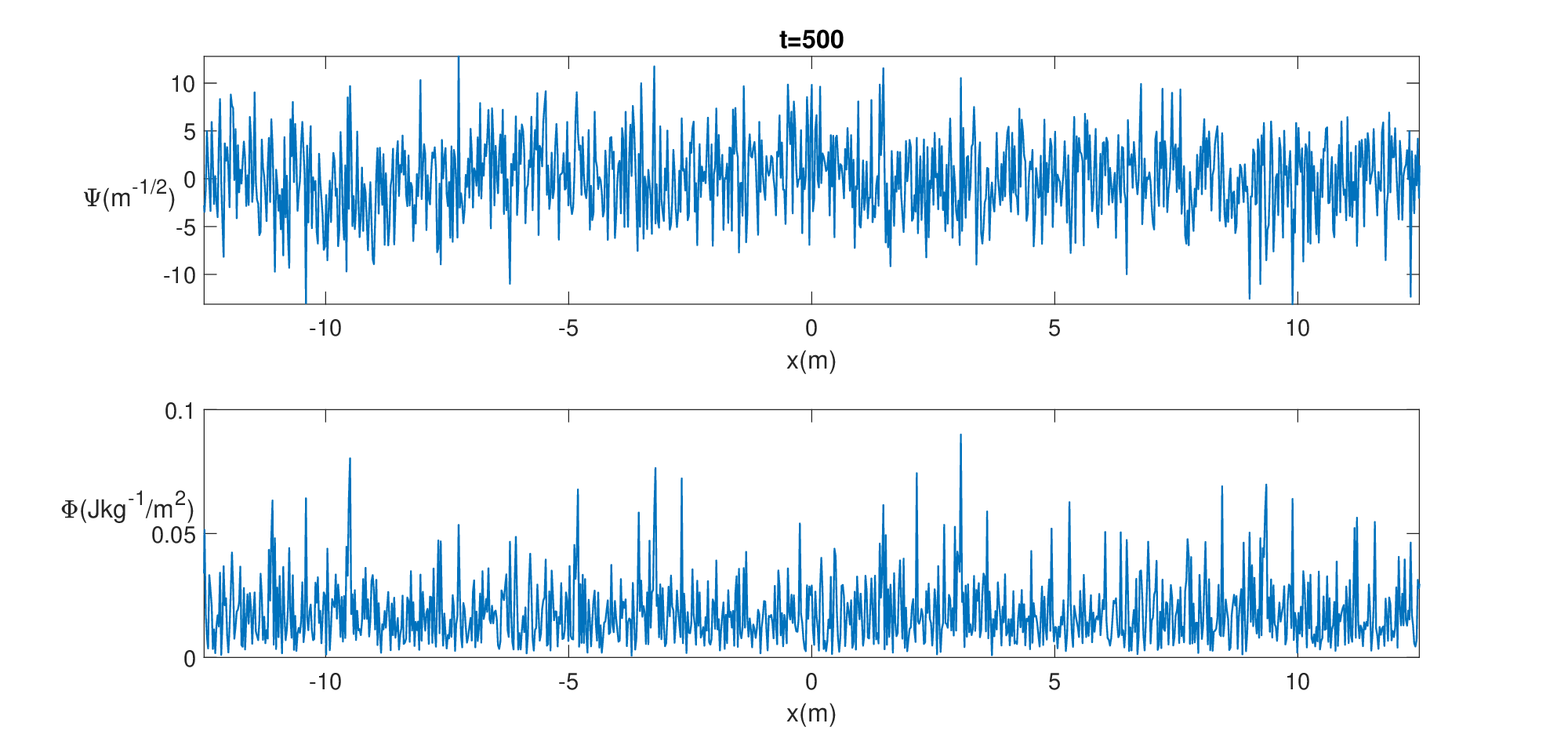}
  \end{center}
\caption{\small The typical chaotic $\Psi$ and $\Phi$ wave fields with rogue waves generated in the frame of the S-N equations for $\beta=0.1$.}
  \label{fig1}
\end{figure}

In order to study rogue quantum gravitational wave dynamics, we perform the numerical solutions of the S-N equations starting from the initial condition $\Psi_0=a_{\Psi i}e^{ix}$ where $a_{\Psi i}$ is a constant. In our simulations, we observe that for the values of $a_{\Psi i} \in [10^{35},10^{80}]$, the wave field initiated by such a monochromatic wave field and excited by the noisy input $\epsilon=\beta a_{\Psi i} v$  given by Eq.(\ref{eq06}), turns into a chaotic one exhibiting many peaks. An example of such a chaotic wave field is shown in Fig. (\ref{fig1}) for $a_{\Psi i}=10^{35}$ and $\beta=0.1$. The peaks in this chaotic wave field have a dynamic nature, they appear and disappear within few time steps. Additionally, they are amplitudes are significantly greater the other waves in the field. Thus, one can possibly name such waves as rogue quantum gravitational waves with an analogy to rogue waves that appear in other fields \cite{Peregrine, Kharif,  Akhmediev2009a, Akhmediev2009b, Akhmediev2011}. The value of the normalization constant for the gravitational field, $a_{\Phi i}$ is found by using the initial condition of $\Psi_0$ in Eq.(\ref{eq02}), which gives $a_{\Phi i}\approx 1.09 \times 10^{-5}$ for the selected initial condition.

\begin{figure}[htb!]
\begin{center}
\hspace*{-0.7cm}
   \includegraphics[width=4.0in]{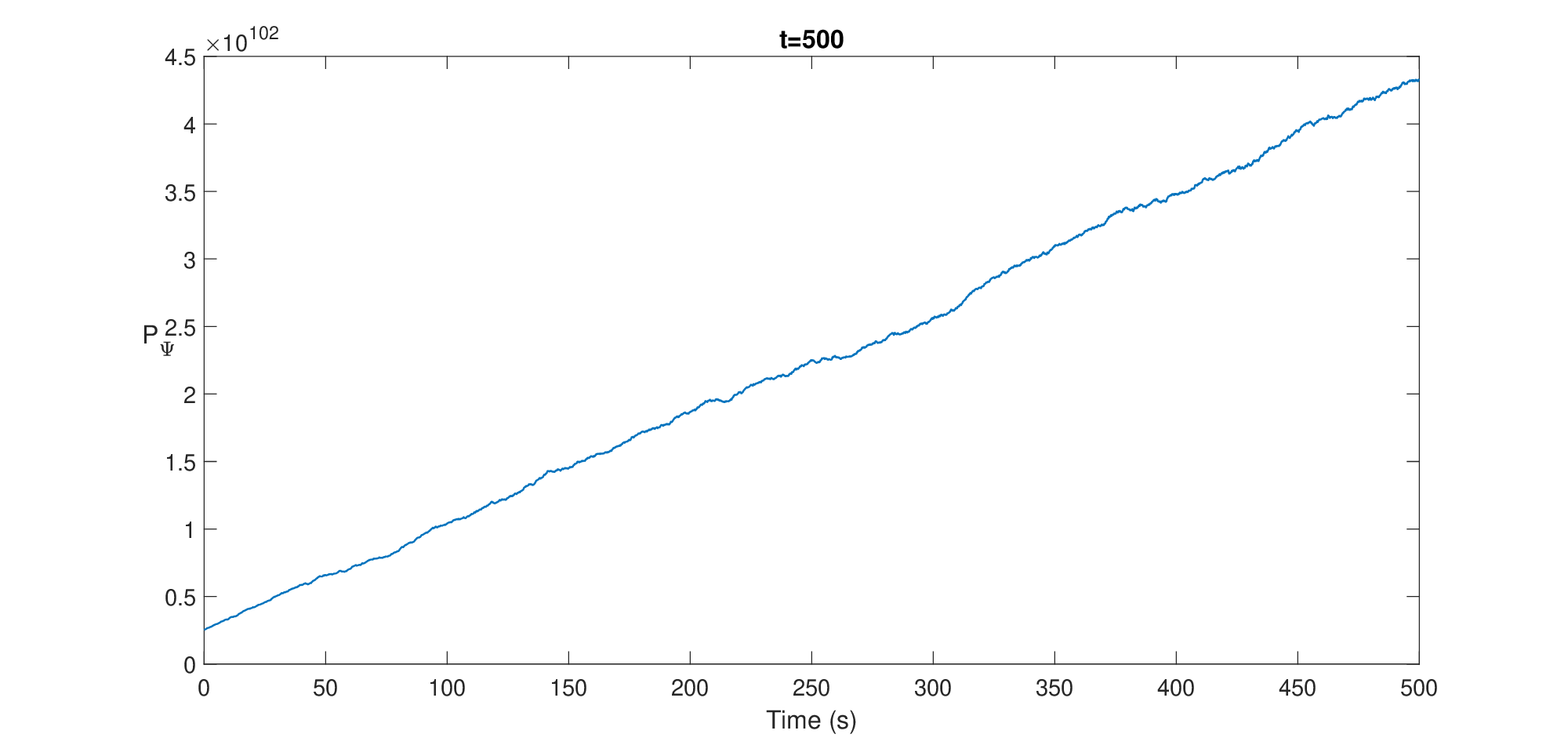}
  \end{center}
\caption{\small Time vs $P_{\Psi}$ graph.}
  \label{fig2}
\end{figure}

\begin{figure}[htb!]
\begin{center}
\hspace*{-0.7cm}
   \includegraphics[width=4.0in]{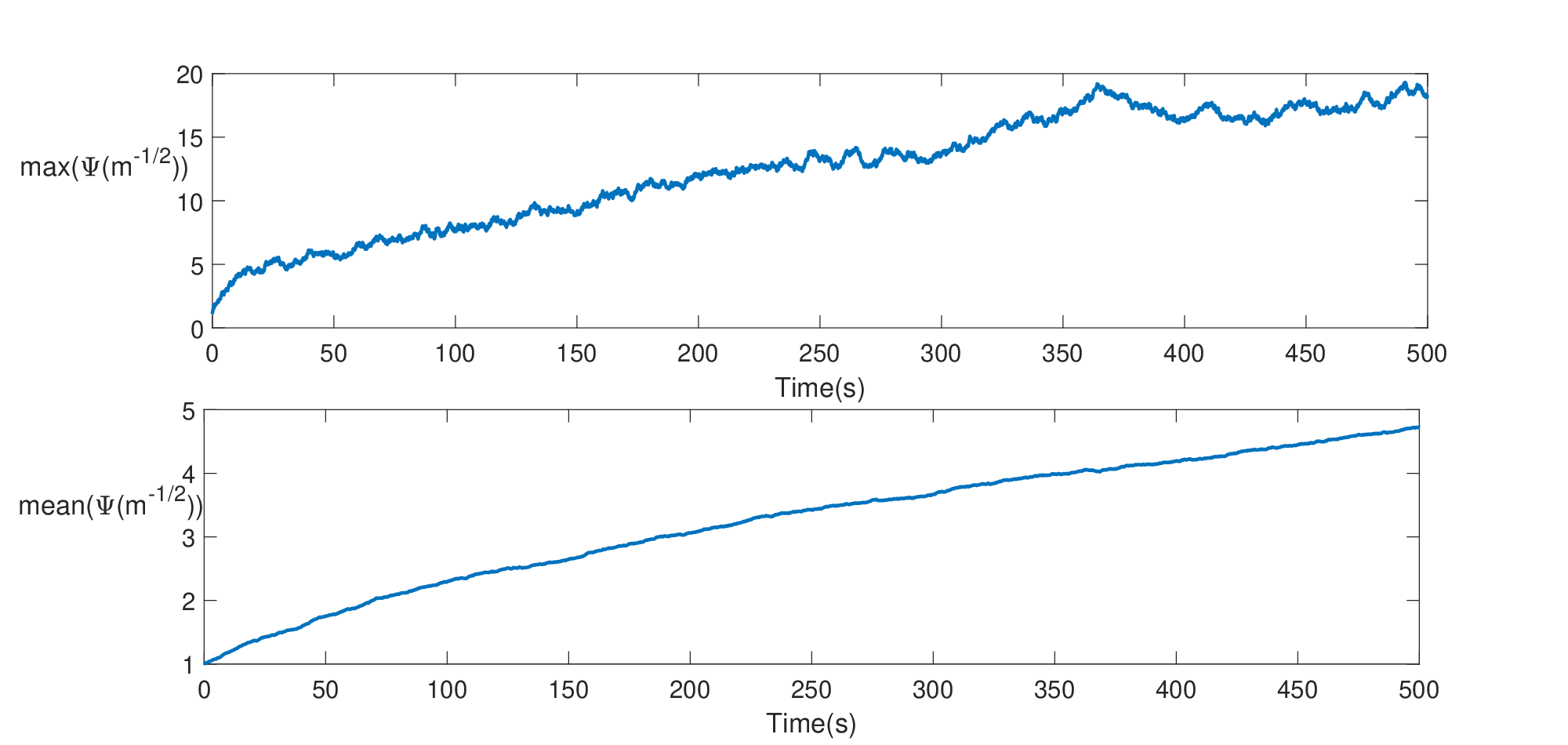}
  \end{center}
\caption{\small Time vs the maximum and mean values of $\Psi$ graph.}
  \label{fig3}
\end{figure}

During the time stepping, we observe that as the amplitudes of peaks of $\Psi$ increases, the amplitudes of peaks of $\Phi$ decreases, indicating a reciprocal relationship. Additionally, we observe that during time stepping the amplitudes of peaks of the wave function $\Psi$ continuously increases, showing a divergent behavior. In order to better illustrate this divergent behavior of the S-N equation system, we depict the time vs power ($P_{\Psi}$) graph in Fig.(\ref{fig2}) and the time vs maximum and mean values of $\Psi$ graph in Fig.(\ref{fig3}). 
 
\begin{figure}[htb!]
\begin{center}
\hspace*{-0.7cm}
   \includegraphics[width=4.0in]{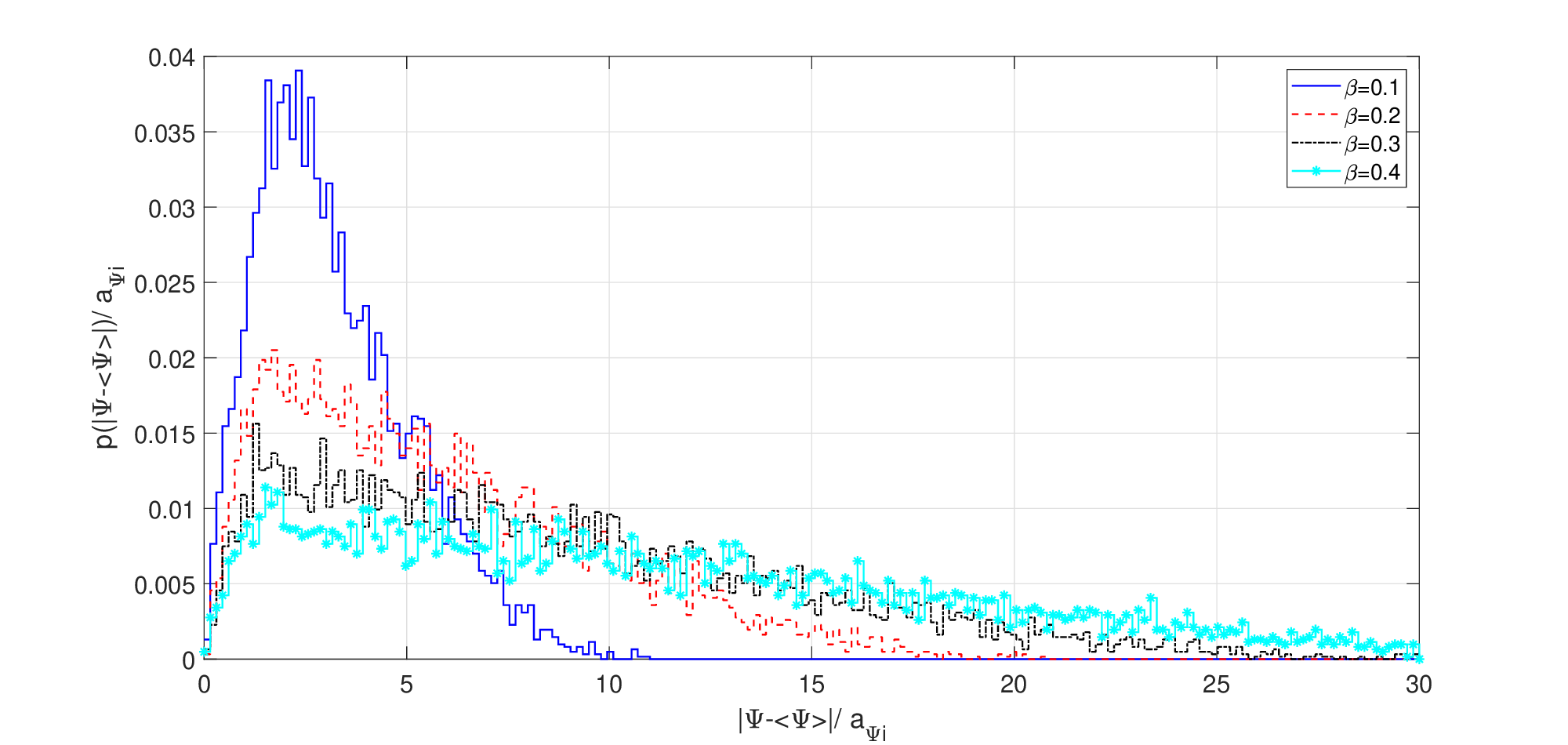}
  \end{center}
\caption{\small Probability distributions of wave function ($\Psi$) amplitudes for various values of the MI parameter $\beta$.}
  \label{fig4}
\end{figure}

\begin{figure}[htb!]
\begin{center}
\hspace*{-0.7cm}
   \includegraphics[width=4.0in]{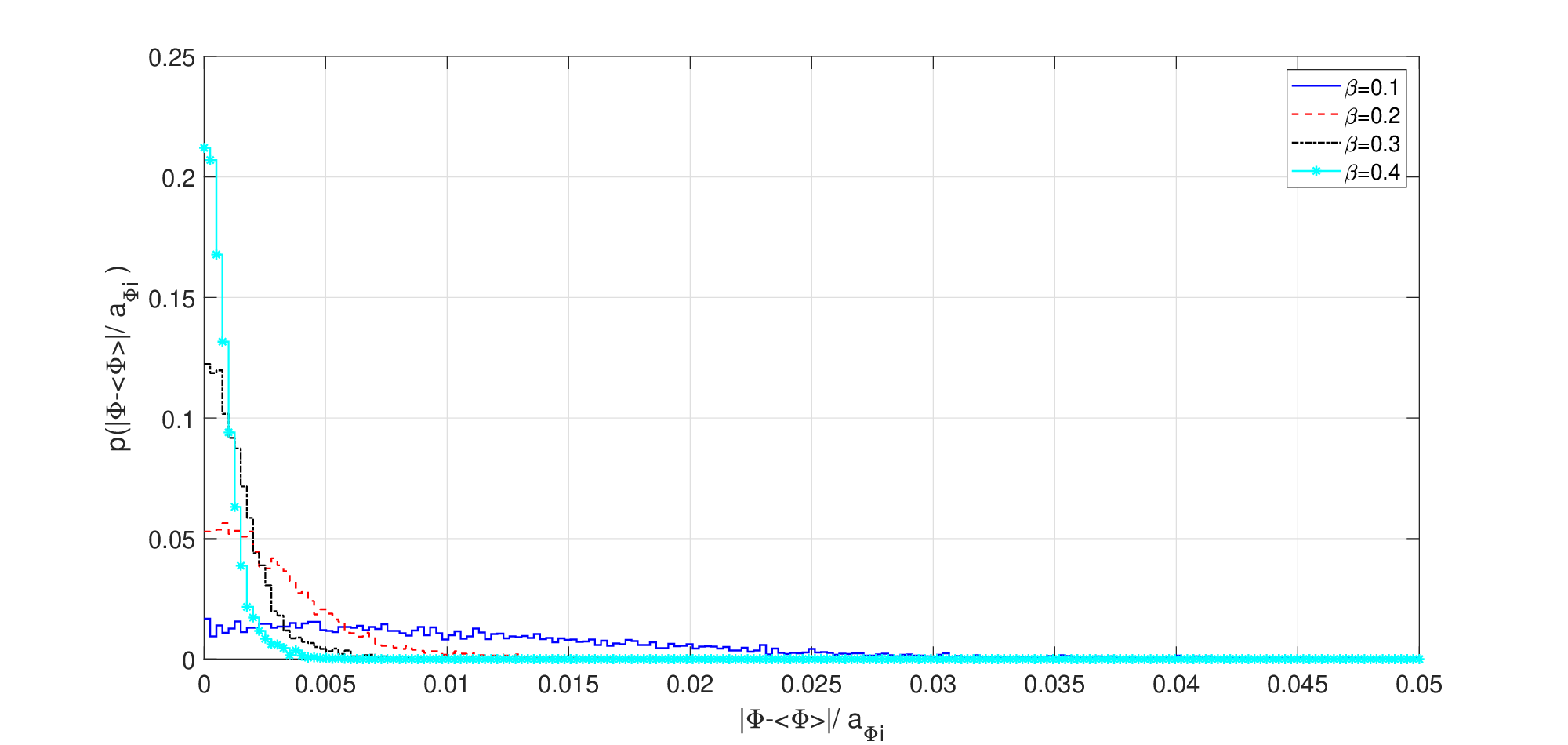}
  \end{center}
\caption{\small Probability distributions of gravitational field ($\Phi$) amplitudes for various values of the MI parameter $\beta$.}
  \label{fig5}
\end{figure}

In Fig.(\ref{fig4}), the probability of wave function amplitude occurrence for various values of $\beta$ is depicted. In order to avoid the effects of the divergence of the S-N equations and to analyze rogue quantum gravitational waves about the mean level, the probabilities are given as the function of the normalized parameter $| \Psi-\left\langle \Psi \right\rangle |/ a_{\Psi i}$. In here, $\left\langle \Psi \right\rangle$ denotes the mean value of the wave function $\Psi$. In order to obtain the probability distributions given in Fig.(\ref{fig4}), the chaotic wave field is recorded at 6 different time steps after an adjustment time of 400s, then these realizations are repeated for 20 times. Each of the curves plotted Fig.(\ref{fig4}) and Fig.(\ref{fig5}) include more than $10^5$ waves. 

As depicted in Fig.(\ref{fig4}), we observe that the probability of rogue quantum gravitational wave occurrence significantly depends and increases with the MI parameter $\beta$. As figure confirms, for $\beta=0.1$ the range of amplitudes for the wave function becomes approximately $|\Psi| \in [0,10]$, however for $\beta=0.4$ this range significantly widens and becomes approximately $|\Psi| \in [0,30]$. The effects of the MI parameter on rogue wave formation probability are studied in other settings in \cite{Soto2014RwSSchaotic, BayPRE1}, which shows a similar tendency. However, the range of amplitudes of the wave function remain mainly in the interval of $|\Psi| \in [0,5]$ in the optical and hydrodynamic media when nonlinear Schrödinger equation is used to model them \cite{Soto2014RwSSchaotic, BayPRE1}. The probability distribution functions depicted in Fig.(\ref{fig4}) closely follows a Rayleigh distribution, with some deviations at the large $|\Psi|$ values, as expected.

In Fig.(\ref{fig5}), we similarly depict the probability of gravitational wave field amplitude occurrence for various values of $\beta$. The probability distributions in this figure are given as the function of the normalized parameter $| \Phi-\left\langle \Phi \right\rangle |/ a_{\Phi i}$ where $\left\langle \Phi \right\rangle$ denotes the mean of the gravitational wave field, $\Phi$. In the frame of the S-N equations, we observe that for the larger wave function amplitudes, the gravitational wave field amplitudes get smaller. As a result of the divergence of the S-N equations, we observe in Fig.(\ref{fig5}) that the probability distributions of the gravitational wave field amplitude move leftwards for larger $\beta$ values. Using the same procedure and the same time steps to get the statistics of gravitational wave field amplitude occurrences as before, we observe that for $\beta=0.1$ the amplitudes lie in the interval of approximately $|\Phi| \in [0,0.05]$, and for $\beta=0.4$ they lie in the interval of approximately $|\Phi| \in [0,0.01]$. Although these amplitudes are much smaller compared to the amplitudes of $\Psi$, the ones which dramatically differ from the significant amplitude of the chaotic wave field at a given time can be named as rogue quantum gravitational waves. Although a consensus is needed to define the rogue gravitational wave, `an unexpected wave having an amplitude at least twice of the significant amplitude at a given time' can be used as a possible definition following the analogous studies in optics, hydrodynamics and Bose-Einstein condensates \cite{Kharif,  Akhmediev2009a, Akhmediev2009b, Akhmediev2011, BayPRE1}. Our results indicate that such waves do exist in quantum gravitational fields, having a wide amplitude range satisfying this criteria.

\begin{figure}[t!]
	\centerline{\includegraphics[width=1.2\columnwidth]{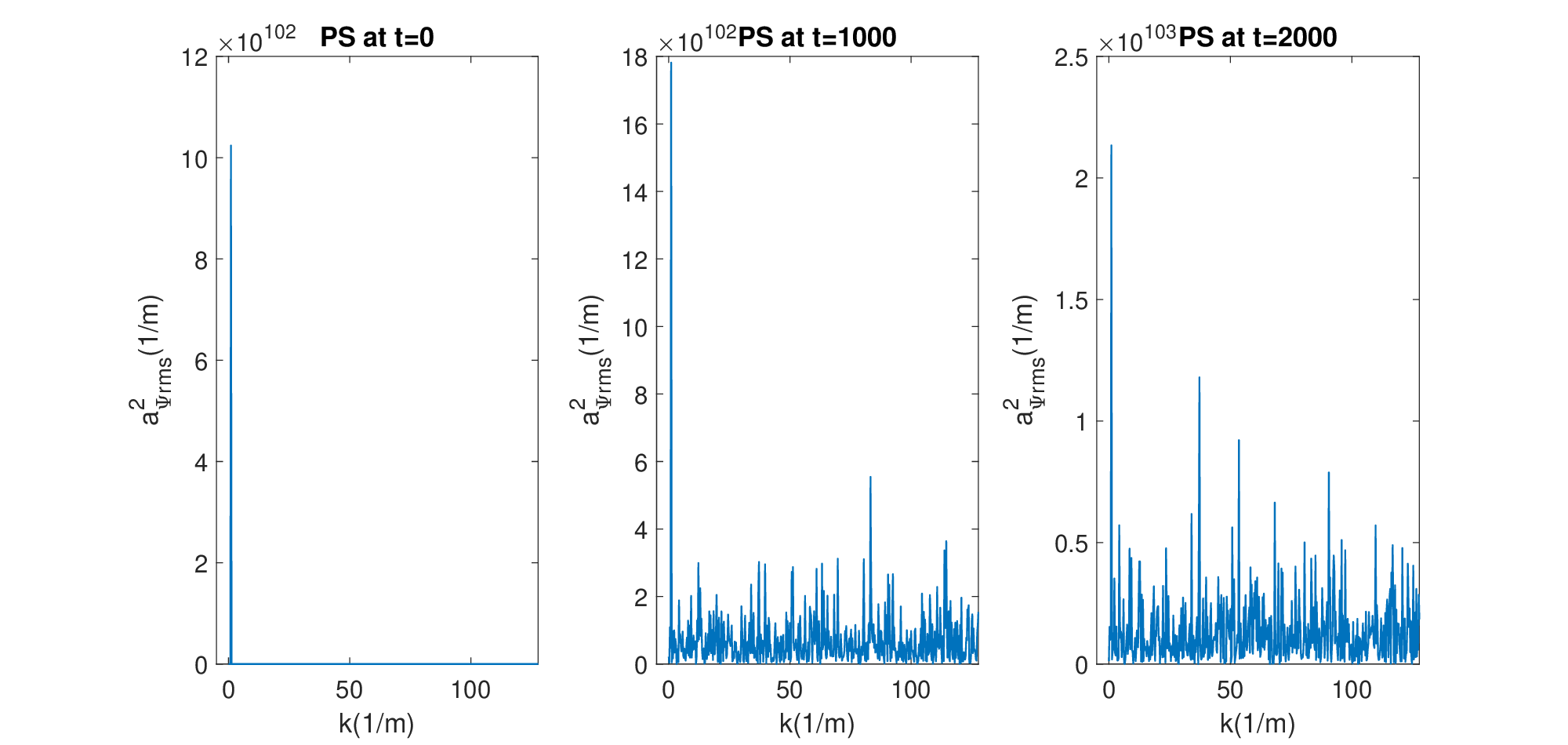}}
	\caption{Temporal evolution of the power spectrum (PS) of the wavefunction $\Psi$. PS at: (a) t=0, (b) t=1000s and (c) t=2000s.}\label{fig:fig6}
\end{figure}

Lastly, we analyze the temporal evolution of the power spectra of the rogue waves generated in the frame of the S-N equations. With this motivation, we depict the one sided power spectra of the wave function $\Psi$ in Figure~\ref{fig:fig6}  for three different times of $t=0s$, $t=1000s$ and $t=2000s$. The power spectra in Figure~\ref{fig:fig6} are calculated by implementing FFT routines and squaring and normalizing the spectral amplitude. As these results confirm, the modulation instability triggered by noise imposed on the sinusoidal wave turns the monochromatic quantum gravitational wave field into a chaotic one and causes energy leakage to higher harmonics in the spectra. This appears as a widening of the spectra, which is also known as supercontinuum generation. It is possible to recognize from Figure~\ref{fig:fig6} for longer temporal evolutions amplitude of transient waves about the mean level grows and becomes more significant. In our simulations, we observe that in the nonlinear regime the supercontinuum generation takes place due to MI and thus wavelengths of the perturbations are not confined by a certain bound such as the one described by the Jeans wave number. It is possible to use these spectral features for the early predictions of rogue quantum gravitational waves before they appear in the physical wavefield \cite{akhmediev2011early, bayindir2016early}.

Rogue quantum gravitational waves can also be studied in alternative ways and within the frame of different quantum gravity models. One of the possible ways of avoiding the divergence of the S-N equations is to introduce a complex gravitational potential \cite{Diosi07, Diosi09}. Thus, the nondivergent dynamics of rogue quantum gravitational waves can be studied in the frame of the S-N equations having a complex gravitational potential proposed and numerically discussed in \cite{Diosi09}. One other possibility is to use the Schrödinger-Newton equation
\begin{equation}
i \hbar \frac{\partial \psi ({\bf x},t)}{\partial t}=-\frac{\hbar ^2 }{2 m}\nabla^2 \psi({\bf x},t)+ \frac{1}{2}m \omega_g^2 \left|{\bf x}- \left\langle  {\bf x} \right\rangle \right|^2 \psi ({\bf x},t)
\label{eq10}
\end{equation}
proposed in \cite{Diosi07} for the center of the mass of a rigid ball. In here, $m$ is the mass of the ball and $\omega_g$ is a certain gravitational frequency of the self-interacting bulk matter \cite{Diosi07}. It is possible to extend this equation to a nonlinear Schrödinger-Newton equation by adding a nonlinear term of $+ \alpha |\psi|^2 \psi$, where the coefficient $\alpha$ show the strength of the nonlinear wave field. It is well-known that these type of nonlinear Schrödinger equations can be derived from the Maxwell's equation using perturbation techniques. It is obvious that, in the frame of such a nonlinear Schrödinger-Newton equation, the MI would trigger formation of rogue quantum gravitational waves.

Our study can shed light upon many different future research directions on quantum gravity. The effects of rogue quantum gravitational waves on particle interactions, quantum vibrations and their interaction with the gravitational field, the wave function collapse and sending coded information in space-time using the gravity may emerge as important fields. With these possible research directions of the near future, the analysis, early detection and efficient sensing of such rogue quantum gravitational waves may turn out to be another important research area. This area would bring many possible applications alongside, including but are not limited to supercontinuum generation, Fourier spectral and wavelet analysis \cite{BayPLA} of rogue quantum gravitational waves and their prediction by using techniques such as the Kalman filtering and deep learning. The analysis on the existence of rogue quantum waves and their study can also be extended to theory of general relativity discussed in \cite{Maggiore}. With this motivation, the Einstein field equations under the effect of noise can be studied, thus gravitational rogue waves may also be hypothesized and observed in the field of general relativity \cite{Maggiore}. With the development of gravitational wave records, the behavior of peaks of the gravitational waves observed in LIGO data can be better understood.

\section{Conclusion}

In this paper, we have analyzed the existence and properties of rogue quantum gravitational waves. With this motivation, we numerically solved the Schrödinger-Newton equations using a spectral scheme with a $4^{th}$ order Runge-Kutta time integrator. We showed that modulation instability turns the monochromatic wave fields into chaotic ones, thus triggers the formation of the unexpected large quantum gravitational waves, which can be named as rogue quantum gravitational waves. We have discussed the characteristics and probabilities such rogue quantum gravitational waves in the frame of the Schrödinger-Netwon equations, however we have also proposed alternative ways of studying such a wave phenomena. Our procedure and findings can lead to many important studies on quantum gravity and may be extended to study rogue gravitational waves in the frame of the Einstein field equations.

\end{document}